\documentclass[%
 reprint,
superscriptaddress,
 amsmath,amssymb,
 aps,
 prl,
]{revtex4-2}

\usepackage{graphicx}
\usepackage{dcolumn}
\usepackage{bm}
\usepackage{xcolor}
\usepackage{url}

\begin{document}

\preprint{APS/123-QED}

\title{Realization of Anomalous Floquet Insulators in Strongly-Coupled Nanophotonic Lattices}

\author{Shirin Afzal}
\email{safzal1@ualberta.ca}
\author{Tyler J. Zimmerling}%
\author{Yang Ren}%
\author{David Perron}%
\author{Vien Van}%
\affiliation{Department of Electrical and Computer Engineering, University of Alberta, Edmonton, AB, T6G 2V4,
Canada}


\begin{abstract}
We experimentally realized  Floquet topological photonic insulators using a square lattice of direct-coupled octagonal resonators.  Unlike previously reported topological insulator systems based on microring lattices, the nontrivial topological behaviors of our system arise directly from the periodic evolution of light around each octagon to emulate a periodically-driven system.  By exploiting asynchronism in the evanescent coupling between adjacent octagonal resonators, we could achieve strong and asymmetric couplings in each unit cell, which are necessary for observing Anomalous Floquet Insulator behaviors.  Direct imaging of scattered light from fabricated samples confirmed the existence of chiral edge states as predicted by the topological phase map of the lattice.  In addition, by exploiting the frequency dispersion of the coupling coefficients, we could also observe topological phase changes of the lattice from normal insulator to Chern and Floquet insulators. Our lattice thus provides a versatile nanophotonic system for investigating 2D Floquet topological insulators.

\end{abstract}

\maketitle


Topological photonic insulators (TPIs) are artificial materials whose electromagnetic band structures exhibit nontrivial topological properties similar to those of electronic wave functions in solid-state topological insulators \cite{bernevig2006quantum,  hasan2010colloquium}.  These materials have attracted a great deal of interest recently due to their exotic properties, such as the existence of topologically-protected edge modes at the sample boundaries, which could be used to realize robust optical devices and other novel applications \cite{ bahari2017nonreciprocal, bandres2018topological, wang2019topological}. Topological insulator behaviors in bosonic systems were first observed at microwave frequencies by applying an external magnetic field to a gyromagnetic photonic crystal to break the time reversal symmetry \cite{wang2009observation}. Since the effect of the magnetic field is weak at optical frequencies, the first photonic topological insulator was realized by emulating a synthetic gauge field in the form of a coupling phase gradient in a two-dimensional (2D) microring lattice \cite{hafezi2013imaging,mittal2014topologically}. Following these works, there have also been realizations of TPIs in zero net magnetic field by introducing a local gauge field using next-nearest neighbor hoppings \cite{leykam2018reconfigurable,mittal2019photonic} or by breaking the spatial symmetry through a deformation of the unit cell \cite{barik2018topological, shalaev2019robust}. All of these realizations of TPIs are based on static systems with time-independent Hamiltonians whose energy bands are well characterized by the Chern number.
More recently, it was shown that Floquet systems with periodically-varying Hamiltonians can exhibit much richer topological properties than static systems.  In particular, Floquet systems can support not only conventional Chern insulator (CI) \cite{fang2012realizing, lindner2011floquet} but also Anomalous Floquet Insulator (AFI) edge modes in the bandgaps between energy bands with trivial Chern number \cite{kitagawa2010topological, rechtsman2013photonic,  rudner2013anomalous, pasek2014network, nathan2015topological, leykam2016anomalous}.  In addition, Floquet systems are more versatile than static systems since their topological behaviors can be tailored through suitable design of the driving Hamiltonian.  AFIs have been demonstrated at acoustic and microwave frequencies using strongly-coupled ring resonators \cite{peng2016experimental,gao2016probing}, and at optical frequencies using 2D arrays of periodically-coupled waveguides \cite{maczewsky2017observation, mukherjee2017experimental,noh2017experimental, mukherjee2018state, guglielmon2018photonic}. For the photonic AFIs based on waveguide arrays, since many periods are required to observe Floquet topological behaviors, the waveguides must have long lengths, typically in the range of centimeters, making them unsuitable for implementation on an integrated photonics platform.

In this paper we report the first experimental realization of AFI on a nanophotonics platform using a lattice of strongly-coupled octagonal resonators in the Silicon-on-Insulator (SOI) material system.  Our system exploits the periodic evolution of light around each microring to emulate a periodically-varying Hamiltonian \cite{afzal2018topological}.  We note that our Floquet TPI lattice is fundamentally different from the microring lattice recently reported in \cite{leykam2018reconfigurable,mittal2019photonic} in that the latter realizes static Chern insulators by emulating a local gauge flux through a combination of direction-dependent nearest and next-nearest neighbor hoppings via link rings between site resonators.  However, since next-nearest neighbor couplings are usually weak, it is difficult to realize AFI behaviors in these lattices, which requires strong coupling to observe.  In our lattice, we exploit asychronism in the evanescent coupling between neighbor octagons to achieve strong and asymmetric direct couplings in each unit cell, which allows us to observe anomalous Floquet topological effects.  Direct imaging of the scattered light pattern shows clear evidence of the formation of chiral AFI edge modes in the bulk bandgaps, which confirms the nontrivial topological behaviors of these lattices.  In addition, by varying the coupling coefficients between the resonators, we could observe topological phase changes in the lattice and verify the existence of both Floquet and Chern insulators in different regions of the topological phase map.  Our work thus introduces a new, versatile integrated optics platform for investigating Floquet topological behaviors in strongly-coupled 2D systems.

\begin{figure}[ht!]
\centering\includegraphics[width=1\linewidth]{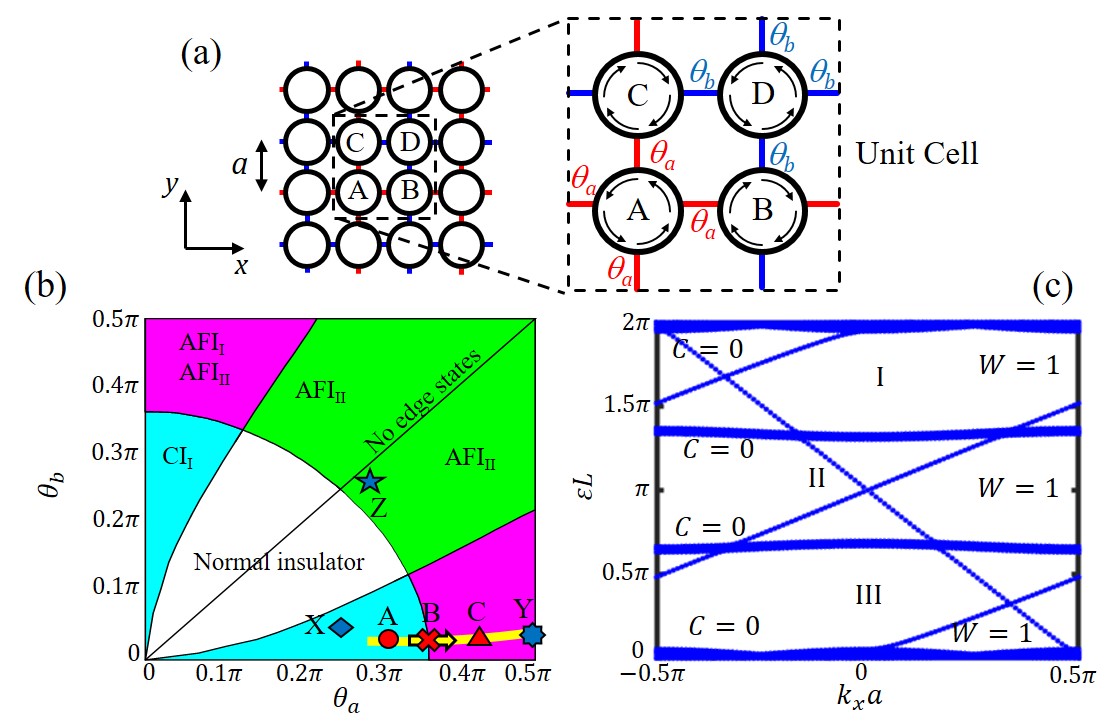}
    \caption{(a) Schematic of a Floquet TPI microring lattice and (b) its topological phase map \cite{afzal2018topological}.  The lattice behaves as a normal insulator except in regions marked by CI\textsubscript{N} or AFI\textsubscript{N}, which denote CI or AFI behavior in bandgap N = \{I, II\} of the lattice. Markers A, B, C correspond to the topological phases of the fabricated lattice at three wavelengths in Fig. 6; X, Y, Z indicate additional fabricated lattices discussed in the Supplemental Material. (c) AFI edge states in the projected quasi-energy band diagram of a semi-infinite lattice with boundaries along the $x$-direction and coupling angles $\theta_a=0.473\pi$, $\theta_b=0.026\pi$.  The Chern number ($C$) of each energy band and winding number ($W$) of each bulk bandgap are also indicated.}
\end{figure}

The topological system in our study is a square lattice of direct-coupled microring resonators with identical resonance frequencies, as depicted in Fig. 1(a).  Each unit cell in the lattice consists of four microring resonators, labelled A, B, C and D, with direct adjacent neighbor couplings represented by coupling angle $\theta$, where $\theta$ is defined such that $\sin^2 \theta$  gives the power coupling coefficient.  Assuming that light in each microring propagates in only one direction, evanescent wave coupling between two neighbor resonators results in a reversal in the propagation direction, or spin flipping.  In this respect, our lattice is also different from the microring lattice in \cite{mittal2019photonic} in that the latter requires off-resonant link rings for coupling between adjacent resonators to emulate a single-spin system, whereas our lattice is much more compact and allows for the natural spin flipping which occurs between direct-coupled microrings.  To realize nontrivial topological behaviors, we allow the coupling strengths between resonator A and its neighbors $(\theta_a)$ to be different from those between resonator D and its neighbors $(\theta_b)$.  The lattice is thus characterized by two coupling angles $\theta_a\neq\theta_b$.  As light circulates around each microring, it interacts periodically with its neighbors via the coupling angles $\theta_a$ and $\theta_b$, with each period equal to one roundtrip $L$ of the microring.  The lattice can thus be regarded as a periodically-driven Floquet system.  The Floquet-Bloch Hamiltonian of the system can be derived by transforming the microring lattice into an equivalent coupled waveguide array, as shown in \cite{afzal2018topological} and briefly summarized in the Supplemental Material.

\begin{figure*}[ht!]
\centering\includegraphics[width=1\linewidth]{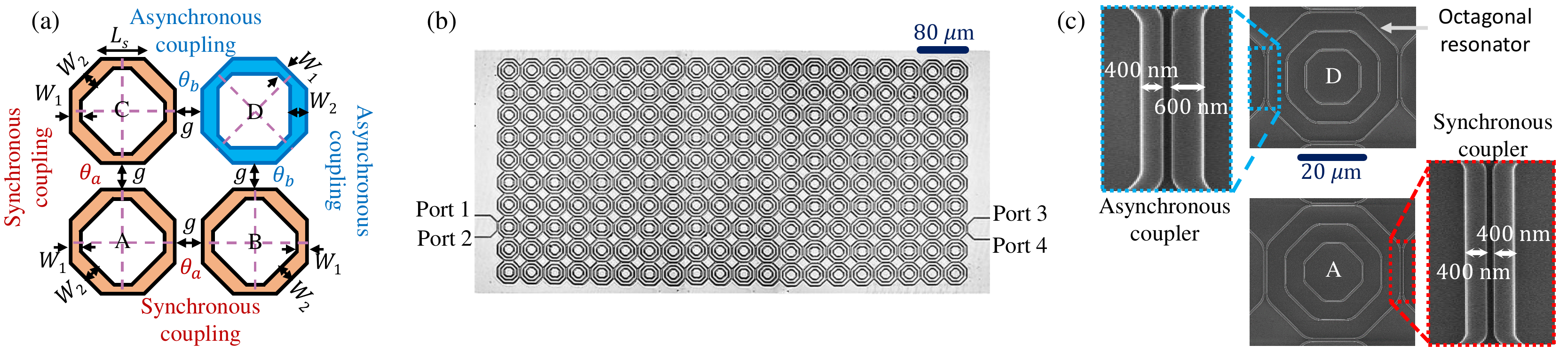}
\caption{(a) Schematic of a unit cell of a Floquet lattice of identical, evanescently-coupled octagon resonators, with octagon D rotated by $45^{\circ}$ with respect to the other three resonators.  (b) Optical microscope image of a $5\times10$ fabricated lattice with input and output waveguides coupled to the left and right boundaries. (c) SEM images of octagonal resonators A and D with zoomed-in images of the synchronous and asynchronous coupling sections (the two smaller octagons inside each octagon are dummy structures).}
\end{figure*}

In general, the band diagram of the quasi-energy $\varepsilon$ of the microring lattice has three bandgaps for every $2\pi$ change in the roundtrip phase of the microrings, \textit{i.e.}, in one free spectral range (FSR) of the resonators.  We label these bandgaps I, II and III, with bandgaps I and III being symmetric about $\varepsilon L = \pi$, as shown in the projected band diagram of a sample lattice in Fig. 1(c).  By computing the Chern number associated with each quasi-energy band and the winding number for each bandgap \cite{rudner2013anomalous}, we can characterize the topological behavior of the lattice in each bandgap as a normal insulator, CI or AFI.  Figure 1(b) shows the topological phase map of the microring lattice in bandgaps I and II as functions of the coupling angles $(\theta_a,\theta_b)$.  Different topological behaviors can be realized by varying the coupling strengths of the lattice.  In particular, AFI behavior is achieved only for strong coupling angles satisfying the approximate relation $\theta_a^2+\theta_b^2 \gtrapprox {\pi}^2/8$ with $\theta_a\neq\theta_b$.  Moreover, the top left and bottom right regions of the map (pink color), where the difference between $\theta_a$ and $\theta_b$ is the greatest, are the only regions in which all three bandgaps are topologically non-trivial, with all exhibiting AFI behavior.  These regions can thus be used to unambiguously demonstrate AFI behavior in the microring lattice. For example, for $\theta_a=0.473\pi$ and $\theta_b=0.026\pi$, which correspond to a lattice realized in this study, the projected band diagram of a sample with 10 unit cells in the $y$-direction and infinite extent in the $x$-direction is shown in Fig. 1(c).  All three bulk bandgaps support edge states of the AFI type, since the winding numbers of the bandgaps are non-zero even though all the energy bands have trivial Chern numbers.



We realized Floquet TPI microring lattices on a Silicon-on-Insulator (SOI) substrate with a $220$ nm-thick silicon layer on a  $2 \mu$m-thick SiO$_2$ layer.  The silicon waveguides were also covered with a $2.2 \mu$m-thick  SiO$_2$ cladding.  For a square lattice of identical resonators, the microrings must be identical and evanescently coupled to their neighbors via identical coupling gaps.  To enable us to realize unequal evanescent coupling strengths for resonators A and D in each unit cell while preserving a square lattice geometry, we used octagonal resonators with the sides having identical lengths but alternating widths $W_1$ and $W_2$, as shown in Fig. 2(a).  Different coupling strengths between adjacent octagons can be achieved by exploiting the difference between synchronous coupling between two waveguides with identical width and asynchronous coupling between waveguides with different widths.  In the lattice, resonators A, B and C are oriented in the same way such that coupling between A and its neighbors B and C occurs synchronously between waveguides of the same width $W_1$.  By rotating resonator D by $45^{\circ}$ with respect to the other three resonators, we could obtain different coupling strengths between microring D and its neighbors due to asynchronous coupling between waveguides of different widths $W_1$ and $W_2$.  In our design, the octagons have sides of length $L_s=16.06 \mu$m, with the corners rounded using 5 $\mu$m-radius arcs to reduce scattering loss.  The octagon waveguide has alternating widths $W_1=400$ nm and $W_2= 600$ nm, which support the fundamental TE mode around the 1600 nm wavelength.  The coupling gap between adjacent octagons is fixed at $g= 225$ nm, yielding $\theta_a=0.473\pi$ for synchronous coupling and $\theta_b=0.026\pi$ for asynchronous coupling around 1620 nm, as obtained from numerical simulations (using the Finite-Difference Time-Domain  solver in Lumerical software \cite{lumerical2018solutions}).  The fabricated lattice consisted of $5\times10$ unit cells, with an input waveguide coupled to resonator A of a unit cell on the left boundary of the lattice for edge mode excitation, and an output waveguide coupled to resonator B of a unit cell on the right boundary for transmission measurements.  The coupling angles between the input/output waveguides and the octagon resonators on the lattice boundary are also set equal to $\theta_a$. Figures 2(b) and (c) show images of the fabricated lattice as well as the synchronous and asynchronous coupling sections.

We characterized the transmission bands of the microring lattice by coupling TE-polarized light to the input waveguide and measuring the transmitted power in the output waveguide (details of the measurement setup are provided in Supplemental Material).  Figure 3(a) shows the normalized power transmission spectrum measured over the $1620{-}1626$ nm wavelength range.  Over one FSR of the microring resonators ($\sim$5 nm), we can identify three bulk bandgaps (I, II and III) separating the passbands.  The high power transmission in all three bulk bandgaps indicate that edge modes are excited in these frequency ranges.  We conclude that these modes must correspond to the AFI edge states which exist in all three bandgaps of the microring lattice, as predicted in the projected band diagram computed for the same lattice in Fig.1(c).  On the other hand, the transmission spectrum in the bulk passbands exhibits multiple dips, which are caused by multiple interference and localized resonances of light propagating through the bulk of the lattice.  For comparison, the simulated transmission spectrum of the lattice computed using the field coupling method in \cite{tsay2011analytic} is shown in Fig. 3(b).  The coupling angles were set at $\theta_a=0.473\pi$ and $\theta_b=0.026\pi$ and a propagation loss of 3 dB/cm was assumed in each octagon resonator.  The effects of $\pm5\%$ uniformly-distributed random variations in the coupling strengths and roundtrip phases in the lattice are also shown by the hatched area in the plot.  The characteristic flat band and high transmission in the bulk bandgaps due to edge modes are clearly visible, in good agreement with the measured spectrum.  The bulk passbands also exhibit transmission dips similar to those observed in the measured spectrum.

\begin{figure}[ht!]
\centering\includegraphics[width=1\linewidth]{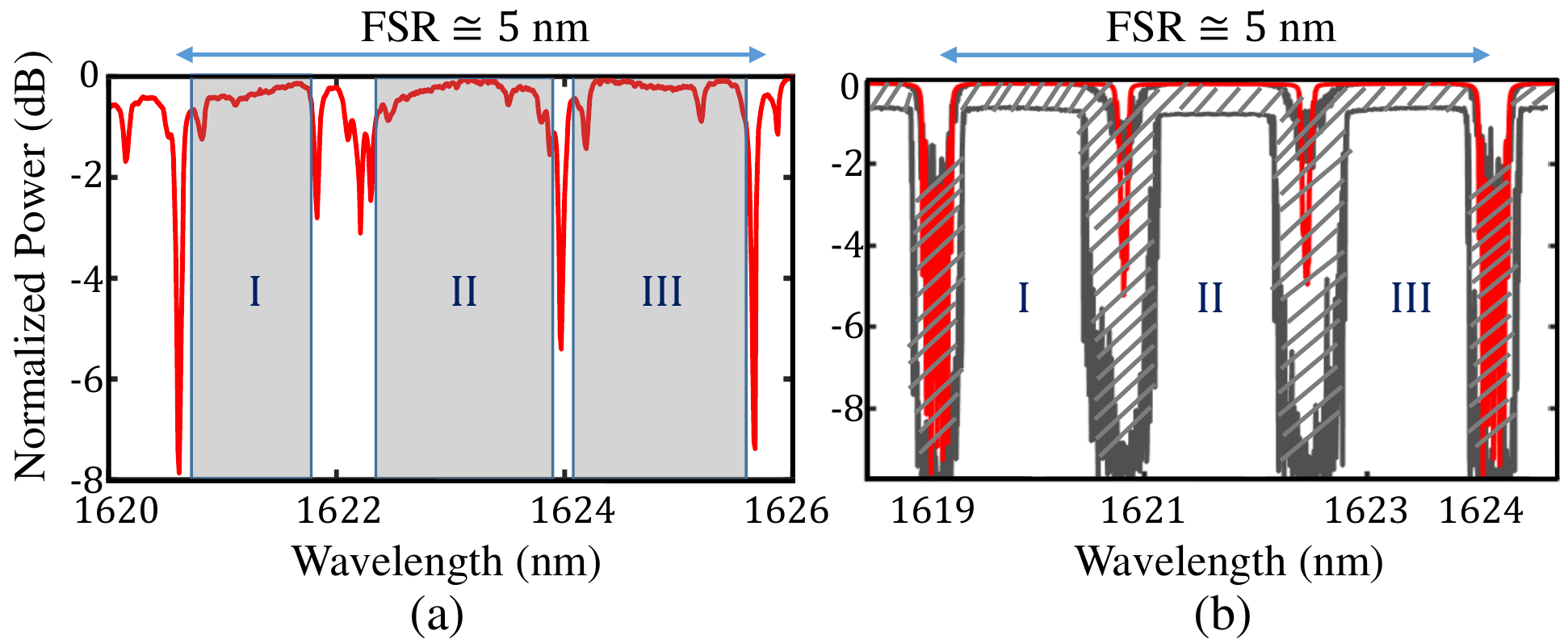}
\caption{(a) Measured and (b) simulated transmission spectra of the TPI microring lattice.  The red line in (b) is the spectrum obtained for an ideal lattice of identical microring resonators with coupling angles $\theta_a=0.473\pi$ and $\theta_b=0.026\pi$.  The hatched area indicates the transmission spectra obtained in the presence of $\pm5\%$ random variations in the coupling strengths and microring roundtrip phases.}

\end{figure}

\begin{figure*}[ht!]
\centering\includegraphics[width=1\linewidth]{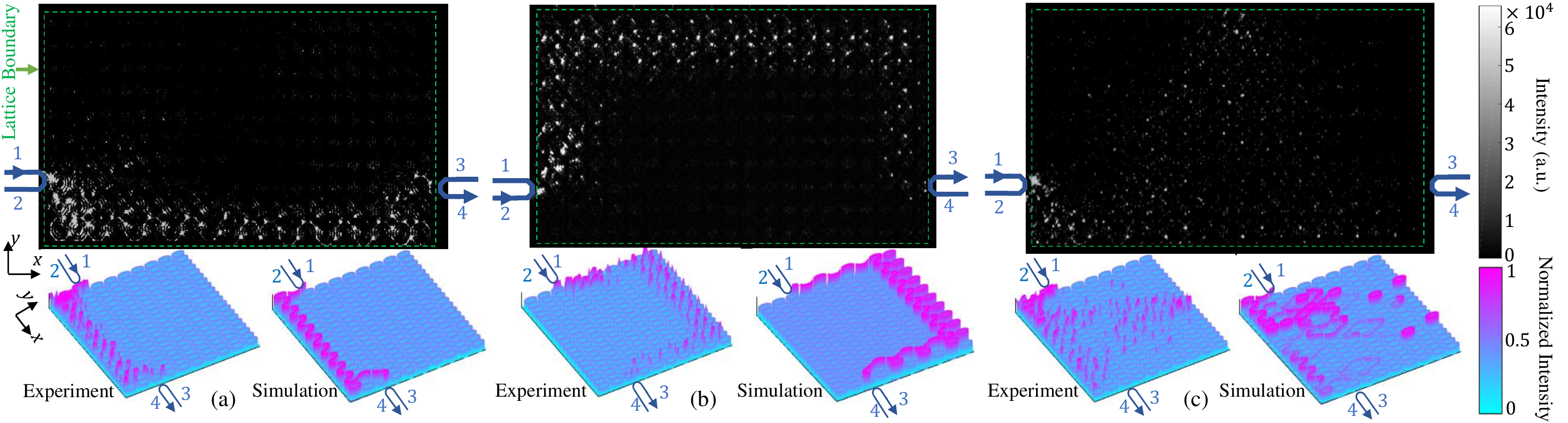}
\caption{(a) and (b) NIR camera images showing formation of chiral AFI edge modes along the bottom edge and top edge, respectively, of the microring lattice when light in a bulk bandgap ($\lambda= 1623$ nm) was injected into Port 1 or Port 2 of the input waveguide.  The lower left plot in each figure shows the map of scattered light intensity constructed from raw camera data; the lower right plot is the simulated light intensity distribution in the lattice.  (c) When the input light was tuned to a wavelength in a transmission band ($\lambda= 1624$ nm), only bulk modes were excited and no edge mode is observed.}
\end{figure*}

To obtain direct proof of AFI edge modes in the bulk bandgaps, we excited the lattice by injecting light at 1623 nm wavelength, which lies in bandgap II, into the input waveguide and imaged the scattered light pattern using a Near Infrared (NIR) camera (details of the imaging system are provided in Supplemental Material).  Figure 4(a) shows the imaged scattered light intensity distribution over the lattice when light was injected into Port 1 of the input waveguide.  Clear evidence of light propagating along the bottom edge of the lattice can be seen, indicating that an AFI edge mode was formed.  The simulated light intensity distribution in the microrings in Fig. 4(a) also shows good agreement with the scattered light intensity map obtained from the camera.  When light was injected into Port 2 of the input waveguide, a counter-propagating edge mode was excited, which propagated along the top edge of the lattice, as seen in Fig. 4(b).  The chiral nature of the edge modes implies that time reversal symmetry is broken between the two pseudo-spin states which exist separately in each microring in the lattice.  We also observed similar AFI edge mode patterns for excitation wavelengths in bandgaps I and III.  By contrast, when we tuned the laser wavelength to 1624 nm, which lies in a bulk passband, only bulk modes were excited and no edge mode was observed.  This can be seen from the NIR camera image in Fig.4(c), which shows that the input light spread out over the lattice instead of being localized along the edge. The simulated light intensity distribution in the lattice at the corresponding wavelength in Fig. 4(c) also confirms this behavior.  



Since the topological behaviors of the microring lattice depend on the coupling angles $\theta_a$ and $\theta_b$, we can observe topological phase changes in the lattice by exploiting the frequency dispersion of the evanescent couplers. Figure 5 shows the simulated coupling angles for the synchronous and asynchronous couplers of the octagon resonators over the wavelength range 1500 – 1630 nm.  The corresponding topological phase of the lattice follows the yellow trajectory in Fig. 1(b).  As the wavelength is tuned from 1500 nm to 1630 nm, the lattice transitions from a normal insulator to an AFI in bandgap II, and from a CI to an AFI in bandgaps I and III.  Figures 6(a)-(c) show the projected band diagrams of a semi-infinite lattice at the three sample points A, B, and C marked on the phase map.  These points correspond to wavelengths $\lambda_A = 1532.5$ nm, $\lambda_B =1546.5$ nm, and $\lambda_C=1593.5$ nm, with coupling angles $(\theta_a;\theta_b)=(0.315\pi; 0.016\pi)$, $(0.355\pi; 0.018\pi)$, and $(0.430\pi; 0.023\pi)$, respectively.  Around wavelength $\lambda_A$, the lattice behaves as a CI in bandgaps I and III and a normal insulator in bandgap II.  At $\lambda_B$, bandgap II closes but the lattice still retains its topological insulator behavior in bandgaps I and III.  Near $\lambda_C$, the lattice supports AFI edge modes in all three bandgaps.

\begin{figure}[ht!]
\centering\includegraphics[width=0.6\linewidth]{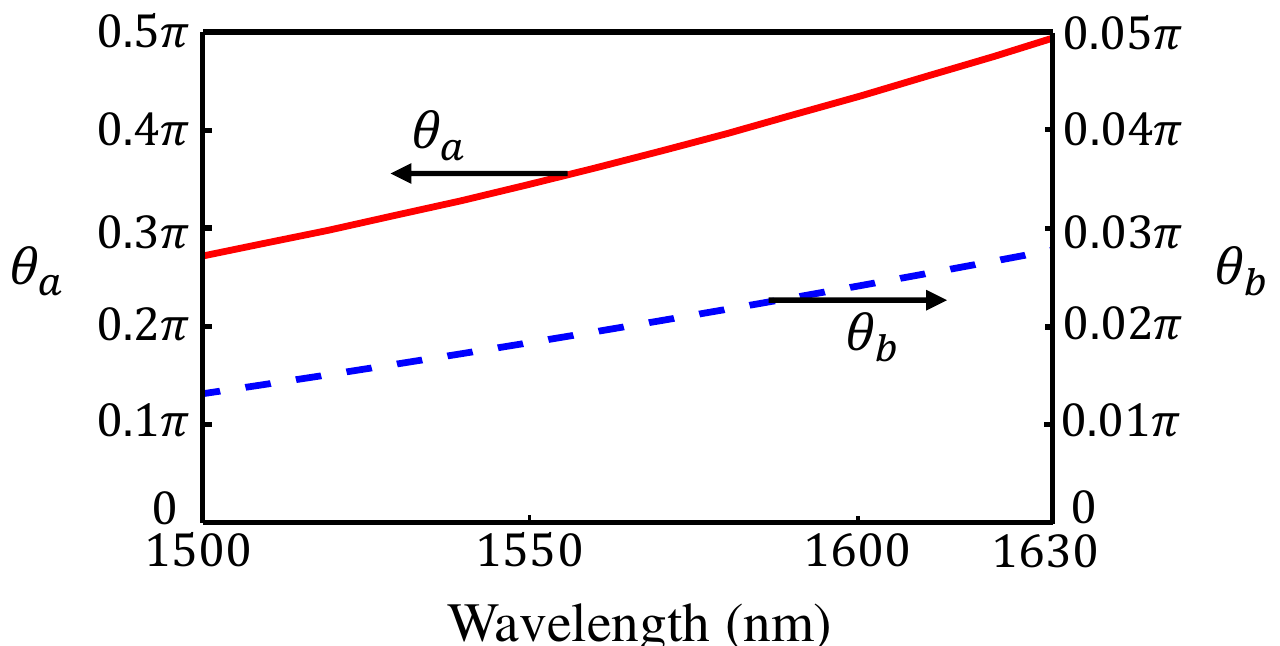}
\caption{Wavelength dependence of the synchronous coupling angle $\theta_a$ and asynchronous coupling angle $\theta_b$ between adjacent octagon resonators.}
\end{figure}

To observe these topological phase changes, we measured the transmission spectra of the microring lattice around the three wavelengths $\lambda_A$, $\lambda_B$, and $\lambda_C$.  The results are shown in Figs. 6(d)-(f).  Close correspondence between the measured transmission spectra and the projected band diagrams can be seen for all three cases.  In particular, high transmission is observed in wavelength ranges corresponding to topologically non-trivial bulk bandgaps where CI or AFI edge modes are expected.  We note that within one FSR of the microring resonators, transmission spectra A and B show only two bulk bandgaps with edge modes (bandgaps I and III) while spectrum C has three distinct bulk bandgaps with edge modes, as predicted by the projected band diagrams.  For spectrum A, the transmission in the center bulk bandgap (bandgap II) is low since the lattice behaves as a normal insulator and thus no edge mode exists.  As the wavelength is tuned from $\lambda_A$ to $\lambda_B$, the center bandgap closes (Fig. 6(e)), although transmission in the output waveguide remains low since light can propagate throughout the lattice and is partially reflected back into the input waveguide.  As the wavelength is further increased to $\lambda_C$ (Fig. 6(f)), the center bandgap opens again but with an important difference in that the transmission in the bandgap is now high, implying the existence of an edge mode.  These transmission spectra thus clearly depict the topological phase change of the lattice in the center bandgap from a normal insulator to an AFI.

\begin{figure}[ht!]
\centering\includegraphics[width=1\linewidth]{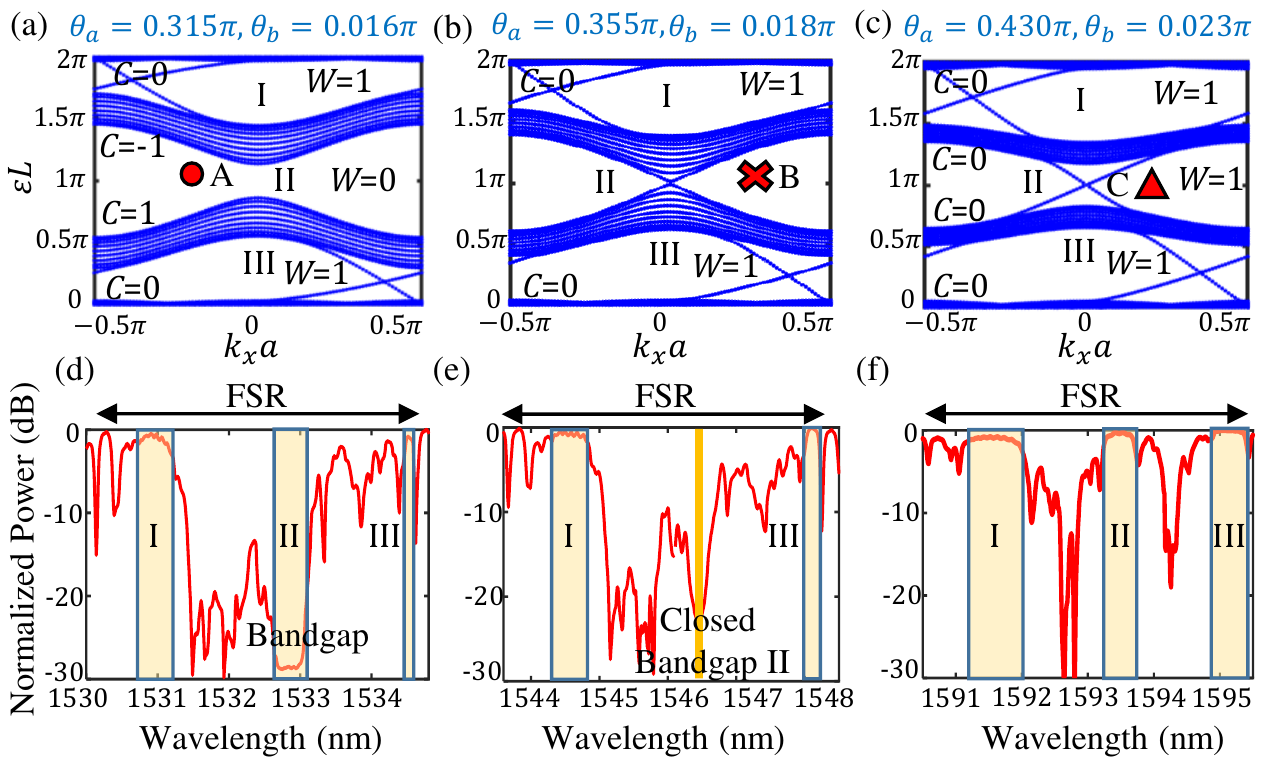}
\caption{Topological phase changes in the octagon lattice due to frequency dispersion in the coupling angles:  (a)-(c) Projected band diagrams of a semi-infinite lattice with 10 unit cells in the $y$-direction, infinite extent in the $x$-direction around (a) $\lambda_A= 1532.5$ nm, (b) $\lambda_B =1546.5$ nm, (c) $\lambda_C =1593.5$ nm. (d)-(f) Measured transmission spectra of the lattice over one FSR centered around the three wavelengths $\lambda_A$, $\lambda_B$, and $\lambda_C$.}
\end{figure}

Direct evidence of these topological phases can also be seen from NIR images of the scattered light distributions at different input wavelengths.  Figure 7(a) shows the scattered light distribution at $\lambda= 1532.80$ nm, which lies in bandgap II of spectrum A where the lattice behaves as a normal insulator. The input light is simply reflected from the lattice in this case.  In Fig. 7(b), the wavelength is tuned to $\lambda= 1534.67$ nm, which lies in bandgap III of spectrum A, where the lattice behaves as a CI.  Clear evidence of an edge mode formed along the bottom edge of the lattice boundary can be seen.  When the wavelength is increased to $\lambda= 1546.50$ nm, which lies in the closed bandgap II of spectrum B, the NIR image in Fig. 7(c) shows light scattered throughout the bulk of the lattice.  Finally, when the wavelength is further increased to $\lambda= 1593.50$ nm, which lies in bandgap II of spectrum C, the lattice behaves as an AFI, as evidenced by the edge mode in Fig. 7(d).

\begin{figure}[ht!]
\centering\includegraphics[width=1\linewidth]{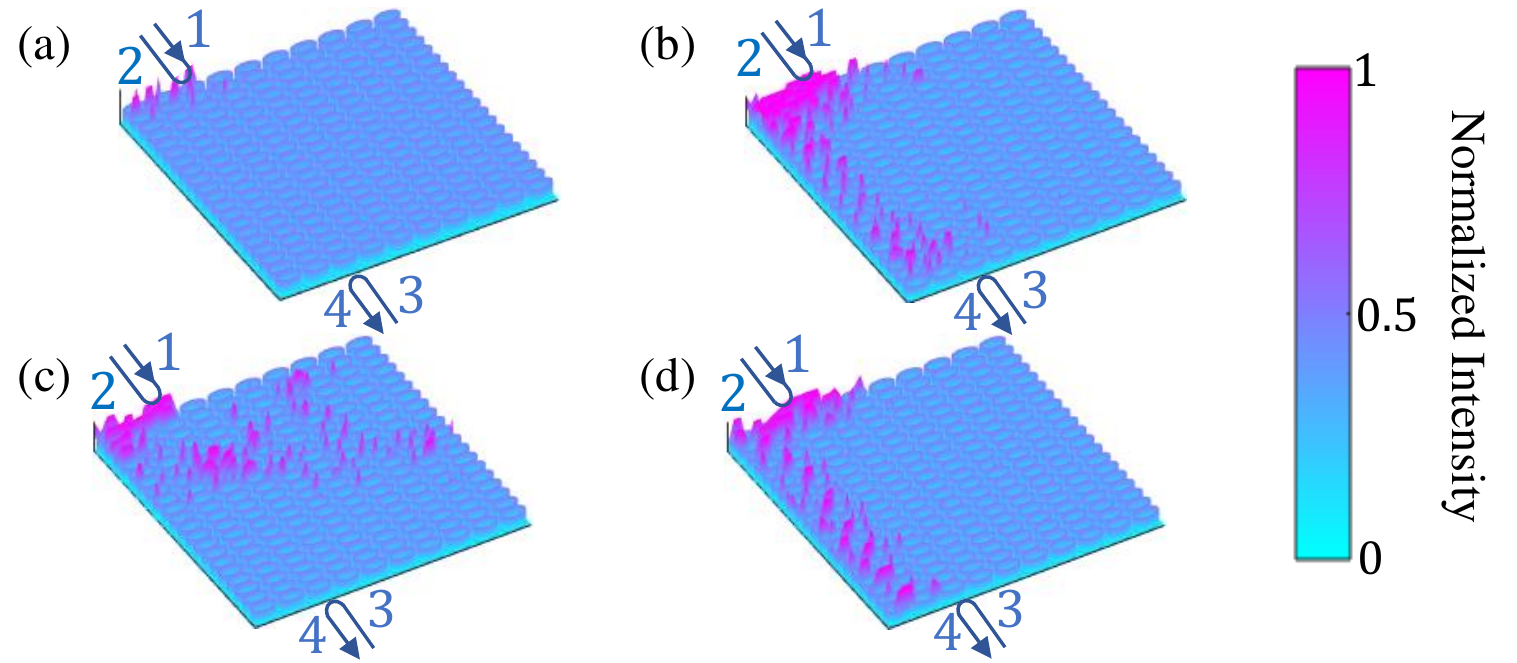}
\caption{Scattered light intensity distributions obtained from NIR camera showing different topological behaviors of the lattice at various input wavelengths: (a) normal insulator at $\lambda= 1532.80$ nm located in topologically-trivial bulk bandgap II, (b) CI edge mode at $\lambda= 1534.67$ nm in bulk bandgap III, (c) bulk modes at $\lambda= 1546.50$ nm in closed bandgap II, (d) AFI edge mode at $\lambda= 1593.50$ nm in reopened bulk bandgap II.}

\end{figure}

We also fabricated octagon lattices with different coupling gap $g$, coupling length $L_s$ and waveguide width $W_2$ to verify the topological behaviors of the lattice in the different regions of the phase map in Fig. 1(b). These samples are marked by X, Y and Z on the map.  NIR images also showed edge modes formed in topologically non-trivial bandgaps of these lattices as predicted by the phase map. These results, which we have included in the Supplemental Material, also provide additional evidence that our Floquet microring lattice behaves as predicted.

In conclusion, we experimentally demonstrated a Floquet TPI based on 2D lattice of strongly-coupled octagonal resonators.  The system emulates a periodically-varying Hamiltonian through the periodic evolution of light around each octagon.  By exploiting asynchronism in the evanescent coupling between waveguides of different widths, we could realize strong and asymmetric direct couplings between adjacent octagon resonators, which allows us to break the time reversal symmetry between pseudo-spin states in the lattice and observe chiral AFI edge modes.  Our lattice also exhibits rich topological behaviors, including normal insulator, CI and AFI, by tuning the coupling angles.  Our work thus introduces a versatile nanophotonic platform for investigating Floquet TPIs and exploring their unique applications. 

\bibliography{ms}

\end{document}



\preprint{APS/123-QED}

\title{Supplemental Material: Realization of Anomalous Floquet Insulators in Strongly-Coupled Nanophotonic Lattices}

\author{Shirin Afzal}
\author{Tyler J. Zimmerling}%
\author{Yang Ren}%
\author{David Perron}%
\author{Vien Van}%
\affiliation{Department of Electrical and Computer Engineering, University of Alberta, Edmonton, AB, T6G 2V4,
Canada}


\maketitle


\section{Derivation of the periodic Hamiltonian of a Floquet microring lattice}

In order to show that a 2D microring lattice emulates a periodically-driven system and derive its Floquet-Bloch Hamiltonian, we first transform the microring lattice into an equivalent array of coupled waveguides, as shown in Fig. S1 \cite{afzal2018topological}.  This is accomplished by "cutting" each microring resonator at the point indicated by the small open circle in Fig. S1(a) and unrolling it to form a straight waveguide.  In this way, we obtain a 2D array of coupled waveguides shown in Fig. S1(b), each with length $L = 2\pi R$, where $R$ is the microring radius. Since light executes many roundtrips in each microring, we can regard the equivalent coupled waveguide array as consisting of a periodic cascade of identical segments of length $L$, so that the fields in the array evolve periodically along the direction of light propagation.  The field at a point in a microring is then the sum of the fields at the corresponding points in all the waveguide array segments.  Using the Coupled Mode Theory, we can write down the equations of motion along length $z$ for the fields $\psi_{m,n}^A, \psi_{m+1,n}^B, \psi_{m,n+1}^C, \psi_{m+1,n+1}^D$ in each unit cell of the waveguide array over one period $L$ as \cite{afzal2018topological}


 \begin{equation*}\label{eq_motion} i\frac{\partial{\psi_{m,n}^A}}{\partial{z}}  = -\beta\psi_{m,n}^A-\kappa_a(1)\psi_{m+1,n}^B-\kappa_a(2)\psi_{m,n+1}^C -\kappa_a(3)\psi_{m-1,n}^B-\kappa_a(4)\psi_{m,n-1}^C \end{equation*}
  \begin{equation*}\label{eq_motion} i\frac{\partial{\psi_{m+1,n}^B}}{\partial{z}}  = -\beta\psi_{m+1,n}^B-\kappa_a(1)\psi_{m,n}^A-\kappa_b(2)\psi_{m+1,n+1}^D -\kappa_a(3)\psi_{m+2,n}^A-\kappa_b(4)\psi_{m+1,n-1}^D \end{equation*}
  \begin{equation*}\label{eq_motion} i\frac{\partial{\psi_{m,n+1}^C}}{\partial{z}}  = -\beta\psi_{m,n+1}^C-\kappa_b(1)\psi_{m+1,n+1}^D-\kappa_a(2)\psi_{m,n}^A -\kappa_b(3)\psi_{m-1,n+1}^D-\kappa_a(4)\psi_{m,n+2}^A \end{equation*}
  \begin{equation}\label{eq_motion} i\frac{\partial{\psi_{m+1,n+1}^D}}{\partial{z}}  = -\beta\psi_{m+1,n+1}^D-\kappa_b(1)\psi_{m,n+1}^C-\kappa_b(2)\psi_{m+1,n}^B -\kappa_b(3)\psi_{m+2,n+1}^C-\kappa_b(4)\psi_{m+1,n+2}^B \end{equation}
 where $\beta$ is the waveguide propagation constant, $\kappa_a(j)=$ sin$\theta_a$ ($\kappa_b(j)=$ sin$\theta_b$) is the coupling strength between microring A (D) and a neighbor in step $j$ of each period. Since the coupled waveguide array is also periodic in $x$ and $y$, we use Bloch\textsc{\char13}s theorem to write $\psi_{m\pm l,n\pm v}^{A,B,C,D}=\psi_{m,n}^{A,B,C,D} {e^{i(\pm lk_x, \pm vk_y) a}} $ where $l$ and $v$ are integers and $a$ is the lattice constant. The equation of motion for a Bloch mode $\textbf{k}$, $|\psi_{m,n}(\textbf{k})\rangle = [\psi^A, \psi^B, \psi^C, \psi^D]_{m,n}^T$, can then be expressed as:  
\begin{equation} i\frac{\partial}{\partial z} |\psi_{m,n}(\textbf{k})\rangle = H_0|\psi_{m,n}(\textbf{k})\rangle + H_{FB}(\textbf{k},z)|\psi_{m,n}(\textbf{k})\rangle  \end{equation}   
where $H_0$ is the unperturbed Hamiltonian and $H_{FB}$
  is the Floquet-Bloch interaction Hamiltonian given by
  \begin{equation} H_{FB}(\textbf{k},z) = - \sum_{j=1}^4 \left( \begin{array}{cc} s_jH_{aa}(j) &(1-s_j)H_{ab}(j) \\ (1-s_j)H_{ab}^\dagger(j)  & s_jH_{bb}(j)
\end{array} \right)  \end{equation}
with
\[ \begin{array}{c}
H_{aa}(j) =  \left( \begin{array}{cc} 0 &\kappa_a(j)e^{i\textbf{b$_j$}.\textbf{k}} \\ \kappa_a(j)e^{-i\textbf{b$_j$}.\textbf{k}}   & 0
\end{array} \right)  \end{array}\] 

\[ \begin{array}{c}
H_{bb}(j) =   \left( \begin{array}{cc} 0 &\kappa_b(j)e^{i\textbf{b$_j$}.\textbf{k}} \\ \kappa_b(j)e^{-i\textbf{b$_j$}.\textbf{k}}   & 0
\end{array} \right)   \end{array}\] 

\[ \begin{array}{c}
H_{ab}(j) =   \left( \begin{array}{cc} \kappa_a(j)e^{i\textbf{b$_j$}.\textbf{k}} &0 \\ 0   & \kappa_b(j)e^{i\textbf{b$_j$}.\textbf{k}}
\end{array} \right)  \end{array}\] 
and  $\textbf{b$_1$}=-\textbf{b$_3$}=(a,0)$, $\textbf{b$_2$}=-\textbf{b$_4$}=(0,a)$, $s_1=s_3=1$, $s_2=s_4=0$. The interaction Hamiltonian is periodic along $z$ with period $L$, $H_{FB}(\textbf{k},z) = H_{FB}(\textbf{k},z+L)$.  The evolution operator of the system is
\begin{equation} 
\label{ev_op} 
U(\textbf{k},z)=T\int_0^z e^{-i H_{FB}(\textbf{k},z')}dz' 
\end{equation}
where $T$ is the time-ordering operator. The Floquet operator $U(\textbf{k},L)$ satisfies the eigenvalue equation
\begin{equation} 
\label{Floqet_op} 
U(\textbf{k},L)|\Psi_n(\textbf{k})\rangle=e^{i\varepsilon_n(\textbf{k})L}|\Psi_n(\textbf{k})\rangle
\end{equation}
where $\varepsilon_n(\textbf{k})$ is the quasi-energy band $n$ of the lattice, which occurs with periodicity of $2\pi/L$. The Chern number of a quasi-energy band $n$ is computed from the eigenstate $n$ of the Floquet operator \cite{rudner2013anomalous},

\begin{equation}
C [P_n] =\frac{1}{2 \pi i} \int_{}^{} d\textbf{k}^2 Tr \{P_n [\partial _{k_x} P_n,\partial _{k_y} P_n]\}
\end{equation}
where $P_n (\textbf{k}) = |\Psi_n(\textbf{k})\rangle  \langle \Psi_n(\textbf{k})| $ is the projector onto the $n^{\textrm{th}}$ eigenstate $|\Psi_n(\textbf{k})\rangle$ of $U(\textbf{k},L)$.  For a Floquet system, the Chern number is insufficient for characterizing its topological behavior \cite{rudner2013anomalous}.  A more relevant topological invariant is the winding number associated with a given energy bandgap.  The winding number depends on the complete evolution history of the system via the evolution operator $U(\textbf{k},z)$ and can be calculated for the microring lattice using the method in \cite{rudner2013anomalous, afzal2018topological}.

\begin{figure}[h!]
\centering\includegraphics[width=0.9\linewidth]{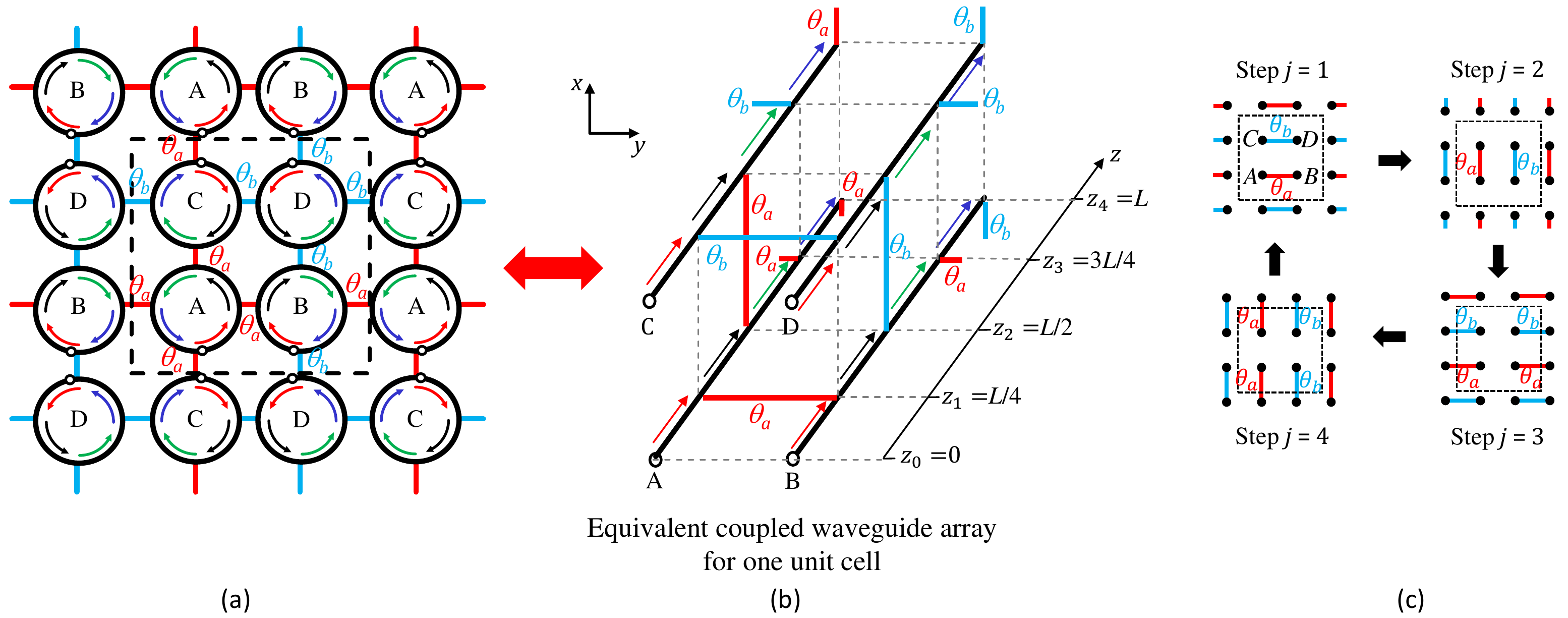}
\caption{Transformation of the microring lattice (a) into an equivalent coupled waveguide array (b) by cutting each microring at the point indicated by the open circle and unrolling it to form a straight waveguide of length $L = 2\pi R$.  In each period $L$, the evolution of fields along the length $z$ of the waveguide array is divided into four steps, each step marked by couplings between pairs of adjacent waveguides via coupling angles $\theta_a$ and $\theta_b$. (c) Hopping sequence in one period showing couplings between pairs of waveguides in each of the 4 steps.}
\end{figure}

The interaction Hamiltonian consists of four coupling steps between adjacent pairs of microring waveguides in each period, as depicted in Fig. S1(c).  The hopping sequence of our microring lattice is more complicated than the hopping sequence realized by the Floquet TPI based on coupled waveguide array in \cite{maczewsky2017observation}.  For our lattice, by having different hopping strengths between sites A and D with their neighbors ($\theta_a \neq \theta_b$), we can guarantee that light in a microring will partially return to its initial location after 3 periods, resulting in localized bulk modes within a topological bandgap.  Figure S2 depicts the driving protocol in the limit $\theta_a \to \pi/2$  $(\kappa_a \to 1)$ and $\theta_b \to 0$ $(\kappa_b \to 0)$. Grey arrows show light hopping in the bulk of the lattice.  Light starting from site A of a unit cell will hop over to site B after the first step due to the strong coupling $\kappa_a$.  In step 2, it remains localized in site B due to the weak coupling $\kappa_b$.  In step 3 it hops over to site A of the right neighbor unit cell.  Continuing tracing the path of light in this manner shows that it will return to its initial position after 3 periods (or 3 microring roundtrips), as shown in the figure. Also indicated in the figure by the purple arrows are the hopping sequences followed by the two chiral edge modes along the bottom and top boundaries of the lattice.

\begin{figure}[ht!]
\centering\includegraphics[width=0.5\linewidth]{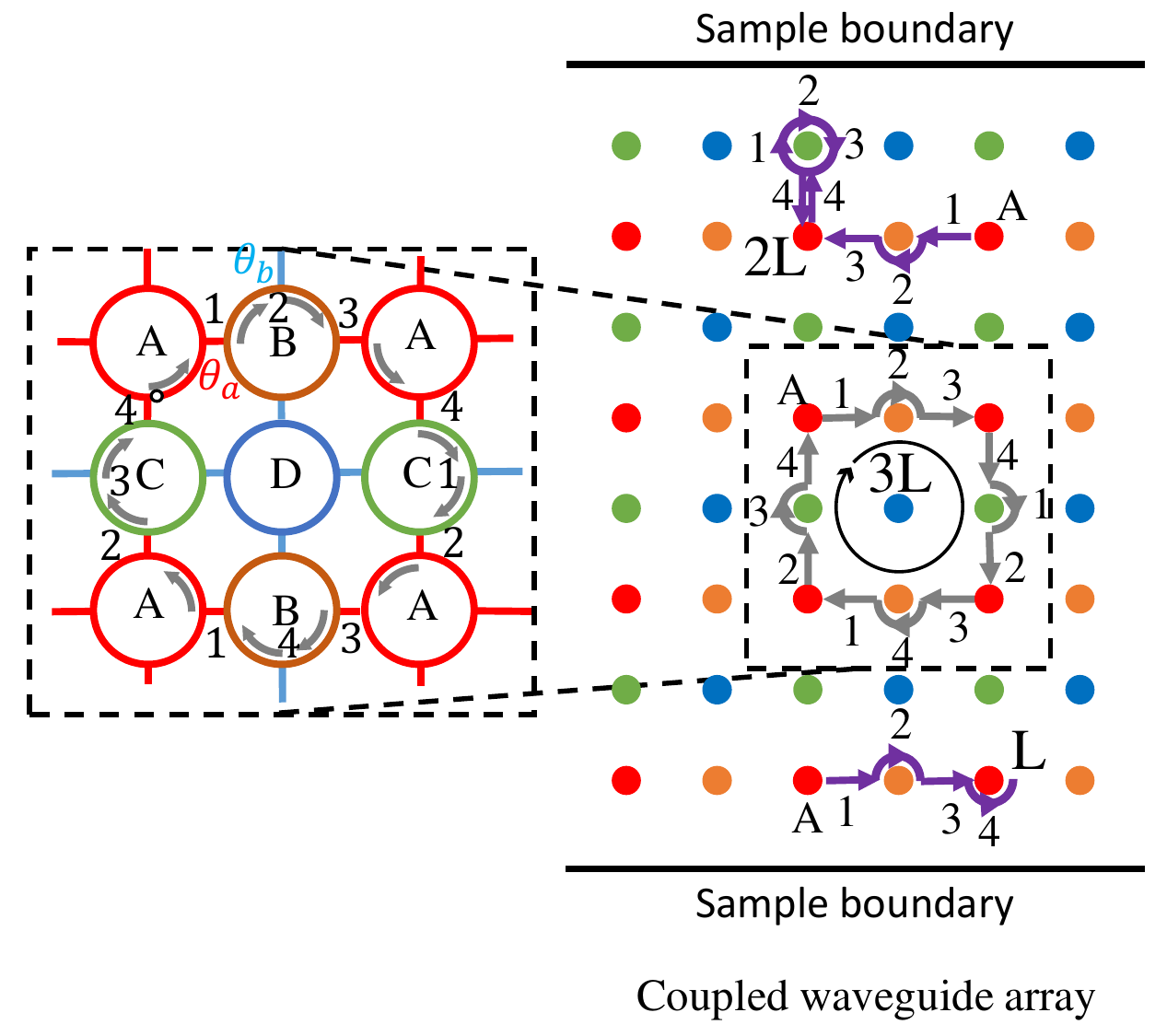}
\caption{Hopping sequence of the Floquet microring lattice in the limit $\theta_a \to \pi/2$ and $\theta_b \to 0$.  Each dot represents a microring waveguide in the equivalent coupled waveguide array picture.  Grey arrows show the path of light in the bulk, which forms a closed loop after 3 periods. The inset in the dashed box shows the path followed by a bulk mode in the microring lattice.  Purple arrows show the hopping sequences of the two chiral edge modes along the bottom and top boundaries of the lattice.}
\end{figure}

\section{Experimental setup for transmission measurement and NIR imaging}

A schematic of the experimental setup used to measure the transmission spectra and perform direct NIR imaging of the microring lattice is shown in Fig. S3.  Light from a tunable laser (Santec TSL-510 $1510{-}1630$ nm) was passed through a fiber polarizer to obtain TE polarization, which was then butt-coupled to the input waveguide via a lensed-tip fiber.  The transmitted light at the output waveguide was collected by another lensed-tip fiber and detected with an InGaAs photodetector and power meter.  To obtained NIR image of the scattered light from the lattice, we used an NIR camera (NX1.7-VS-CL-640) to image the lattice through a 20$\times$ objective lens.  The NIR camera had a 14 bit digital InGaAs image sensor with 640$\times$512 pixels and 15 $\mu$m pixel pitch.  Using the raw data output from the camera, we also constructed a 2D map of the scattered light intensity which was overlaid on a drawing of the lattice for better visualization of the edge modes.
\begin{figure}[ht!]
\centering\includegraphics[width=0.5\linewidth]{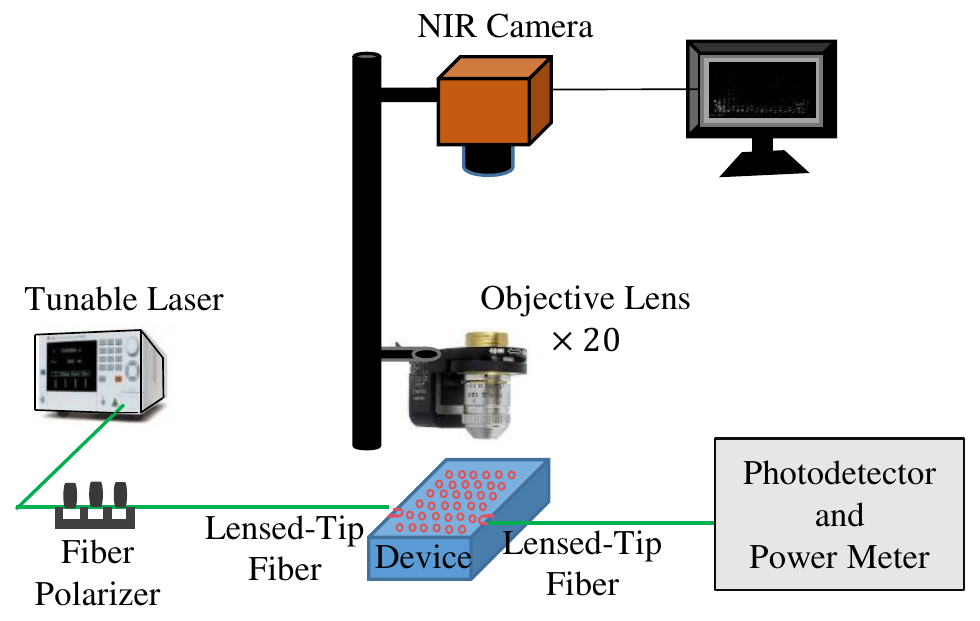}
\caption{Schematic of the experimental setup used to perform transmission measurements and NIR imaging of  scattered light from the microring lattice.  }
\end{figure}

\section{Edge modes of octagon lattices located in other regions of the topological phase map}

In order to confirm the topological behaviors of the coupled octagon lattice in other regions of the phase map in Fig. 1(b) of the main text, we also fabricated three other lattices with different values of coupling angles $\theta_a$ and $\theta_b$ and performed NIR imaging of the edge modes.  Table S1 lists the design parameters of the fabricated lattice samples.  We obtained different coupling angles by varying the coupling gap $g$, coupling length $L_s$ and the octagon waveguide width $W_2$.  For each sample, we also showed in Table S1 the computed coupling angles $\theta_a$ and $\theta_b$ at the wavelength at which we imaged the edge modes.  The location of each sample on the topological phase map is shown in Fig. S4(a).  Sample X consisted of $5\times10$ unit cells, similar to the lattice shown in Fig. 2(b) of the main text.  Samples Y and Z were twice as large with $10\times10$ unit cells, an image of which is shown in Fig. S4(b).    


\begin{table}[h!]
\caption{Design parameters of the fabricated octagon lattice samples}\label{tab1}
\begin{center}
  \centering
  \begin{tabular}{|c|c|c|c|c|c|c|c|c|}
    \hline
    Sample   & $L_s$($\mu$m)    & $W_1$(nm) & $W_2$(nm) & $g$(nm) &  $\theta_a(\pi)$&  $\theta_b(\pi)$ & wavelength& edge mode type\\
 \hline
X  &  $13.14$ & 400 & 600 & 275 &  $0.254$&  $0.031$& 1627 nm & CI  (bandgap I) \\
\hline
Y &  $16.06 $ & 400  & 600 & 200  &  $0.5$&  $0.027$ & 1600 nm  & AFI (bandgap  II)   \\
 \hline
Z &  $16.06 $ & 400 & 410 & 225 &  $0.307$&  $0.277$& 1527 nm & AFI  (bandgap  II)  \\
 \hline
  \end{tabular}
 \end{center} 
\end{table}

\begin{figure}[ht!]
\centering\includegraphics[width=0.8\linewidth]{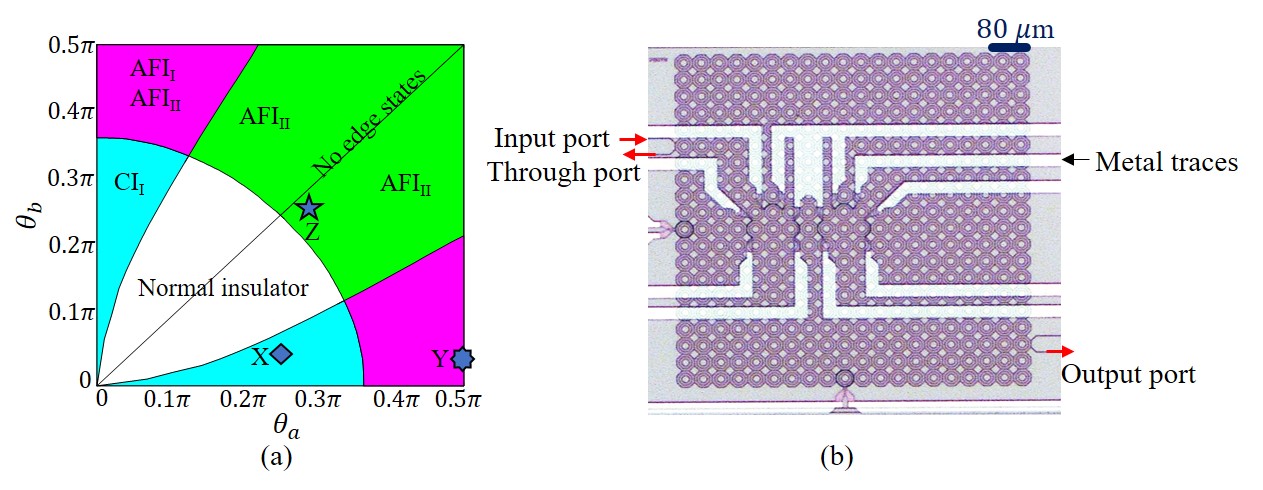}
\caption{(a) Locations of the fabricated octagon lattice samples X, Y, Z in the different regions of the topological phase map. (b) Optical microscope image of sample Z with $10\times10$ unit cells (the metal traces on the lattice were for performing other studies on the lattice).}  
\end{figure}

For each sample we measured the transmission spectrum of the lattice and then tuned the laser wavelength to a bulk bandgap where we expected to observe an edge mode.  The wavelength, bandgap number and type of the edge mode (CI or AFI) are also listed in Table S1.  The NIR images of the scattered light distribution of the samples are shown in Figs. S5(a)-(c).  Figure S5(a) shows CI edge mode in bandgap I of sample X, whose existence is predicted in the phase map.  Figure S5(b) shows AFI edge mode in bandgap II of sample Y.  For this lattice, only one waveguide was coupled to the lattice (on the left boundary) so the excited edge mode is seen to make a complete trip around the perimeter of the sample in the counterclockwise direction before exiting at the through port of the waveguide. The intensity of the scattered light is also seen to decay along the propagation path due to loss.  Figure S5(c) shows an AFI edge mode in bandgap II of sample Z.  For this sample the two coupling angles were only slightly different ($\theta_a=0.307\pi$ and $\theta_b=0.277\pi$).  The edge mode was excited by injecting light through an input waveguide coupled to the left side of the sample.  Light is seen to propagate along the sample boundary in the counterclockwise direction and exit through an output waveguide on the right side of the sample.  We observe that due to the smaller difference between $\theta_a$ and $\theta_b$ for this lattice, the edge mode was not as tightly localized near the sample edge compared to sample Y but penetrated more deeply into the lattice bulk.  These results provide additional evidence that our coupled octagon lattices possess the topological properties as predicted by the phase map.

\begin{figure}[ht!]
\centering\includegraphics[width=0.8\linewidth]{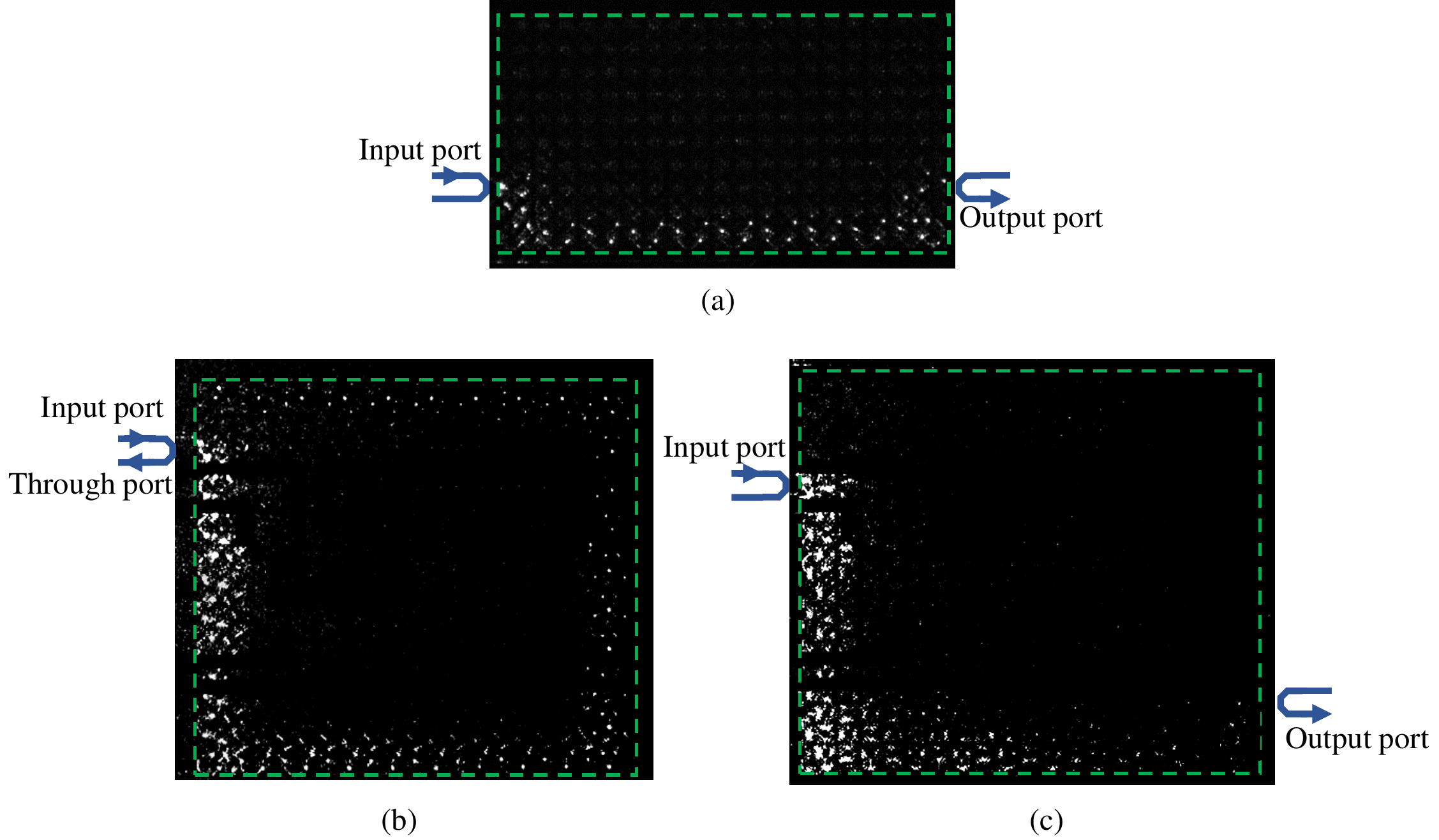}
\caption{NIR images of scattered light distribution showing (a) CI edge mode in bandgap I of sample X, (b) AFI edge mode in bandgap II of sample Y, and (c) AFI edge mode in bandgap II of sample Z.}
\end{figure}



\bibliography{supplement}